\documentclass[%
 reprint,
 amsmath,amssymb,
 aps,
]{revtex4-2}

\usepackage{graphicx}
\usepackage{dcolumn}
\usepackage{bm}
\usepackage{hyperref}
\usepackage{xcolor}

\newcommand{\be}{\begin{equation}}
\newcommand{\ee}{\end{equation}}
\newcommand{\bea}{\begin{eqnarray}}
\newcommand{\eea}{\end{eqnarray}}

\newcommand{\Bl}{\color{black}}

\newcommand{\vecb }{\boldsymbol}
\newcommand{\bs }{\boldsymbol}
\hypersetup{
    colorlinks = true,
    linkcolor  = blue,
    citecolor  = blue,
    urlcolor   = black
}

\begin{document}

\preprint{APS/123-QED}

\title{Orbital Hall effect from the Wannier function approach}

\author{Tapesh Gautam}
\affiliation{Department of Physics \& Astronomy, University of Missouri, Columbia, MO 65211, USA} 

\author{S. Satpathy}
\email{satpathys@missouri.edu}
\affiliation{Department of Physics \& Astronomy, University of Missouri, Columbia, MO 65211, USA} 

\date{\today}

\begin{abstract}

Using the Wannier function approach, we develop a gauge-covariant formalism for the orbital angular momentum (OAM) operator and the orbital Hall conductivity (OHC).
The Wannier function formulation resolves the position-operator ambiguity inherent to Bloch states, providing a transparent and rigorous theory of orbital transport.
Our formalism is validated through a finite-lattice calculation of the Haldane model, where the OAM operator is evaluated exactly 
from its real-space definition.
Our results are also relevant to other orbital  phenomena in solids such as 
the orbital Edelstein effect, orbital torque generation, and the like, where the OAM plays a crucial role. 
\end{abstract}

\maketitle

\section{Introduction}

Orbital transport has garnered considerable interest in the past decade as a fundamental component of the electronic transport phenomena in solids \cite{rev1, rev2, rev3}. 
In particular, the orbital Hall effect (OHE) \cite{Kotani2008, Kotani2009, ohe_go, ohe-SB-SS-1, ohe_tmd_rappoport, ohe-SB-SS-2, choi_exp, Lyalin_exp, Idrobo_exp, ohe_arovas} and the orbital Edelstein effect \cite{oee_annika, oee_annika2, oee_annika3, oee_tapesh, oee_pratik} have emerged as central concepts in orbitronics, describing, respectively, the generation of transverse orbital angular momentum (OAM) currents and the non-equilibrium accumulation of orbital moments under applied electric fields. The injection of an OAM current into a FM material can induce an orbital torque, akin to the spin torque induced by a spin current \cite{Go2019, torque1, Rappoport-torque,  ding_torque}.  In some materials, these orbital responses have been predicted to surpass their spin counterparts \cite{ohe-SB-SS-1, oee_tapesh}. 
The OAM is central to the description of orbital transport phenomena such as the orbital Hall effect, the orbital Edelstein effect, and orbital torque.

Despite this progress, the theoretical description of the  OAM in solids remains subtle due to the ill-defined nature of the position operator for the Bloch functions\cite{blount, resta_1998}.
 Many early theory works used the so called ``atom-centered approximation (ACA)," where the OAM was replaced by the atom-based angular momentum operators. Although this was sufficient in many cases to provide insights into the basic physics, it is nevertheless an unregulated approximation and completely fails for some materials, e.g., graphene, with zero atomic angular momentum \cite{ohe_vignale}. Recently, several authors have described the OAM using the momentum-space approach, but with conflicting results \cite{ohe_vignale, ohe_pezo, ohe_rappoport}. 

In this work, we use the Wannier function approach to treat the OAM operator, which leads to a gauge-invariant theory of the orbital Hall effect within the Kubo linear response framework. 
{\Bl 
In contrast to the Bloch representation, the position operator is well defined in the Wannier representation, which enables a transparent derivation of the OAM,
thus eliminating the ambiguities in the treatment of the position operator 
inherent in the reciprocal space.
}
We validate our results by computing the OAM as well as the OHC for a finite-lattice Haldane model\cite{Haldane1988}, where we evaluate 
the OAM exactly from its direct definition in real space. 
The agreement between the two establishes our results.

\section{OAM operator}

The key difficulty  in the computation of the orbital Hall conductivity (OHC) is the calculation of the OAM matrix elements $\langle \psi_{n \vecb k}| \vecb r \times \vecb v | \psi_{n'\vecb k} \rangle$  between the Bloch states, where $\vecb r$ is the position operator and
    $\vecb{v} =  - i\hbar^{-1}  [ \vecb{r},\mathcal{H} ]$
is the velocity operator. We use the symmetric form\cite{ohe_vignale}  
of the orbital angular momentum operator
\be
\vecb L =  \frac{m_e}{2} \  (\vecb{r}   \times \vecb{v} - 
 \vecb{v} \times \vecb r),
 \label{eq4}
\ee
where $m_e$ is the electron mass. The orbital magnetic moment\cite{Note2} is $\vecb m = -(g\mu_B/\hbar) \vecb L$. 
The difficulty of the calculation of OAM stems from the fact that the position operator is ill-defined for the Bloch states. Indeed, while the matrix element remains invariant to the shift of the origin by a lattice translation vector, in general, the result is different if $\vecb r$  is shifted by a fractional translation within the unit cell.
This is an important point that was analyzed in the seminal work by Niu and coworkers\cite{niu_1996, niu_1999, niu_2005_center}, where by considering the energetics of a Bloch wave packet within a semiclassical picture, it was shown that the origin 
should be shifted by  
$\vecb r_{n \vecb k} = i\langle u_{n \vecb k} | \partial_{\vecb k} u_{n \vecb k} \rangle \equiv
\vecb{{\mathcal A}}_{n\vecb k} $, 
the Berry connection, so that the diagonal OAM matrix element is   $\langle \psi_{n \vecb k}| (\vecb r-\vecb r_{n \vecb k}) \times \vecb v | \psi_{n\vecb k} \rangle
= \langle u_{n \vecb k}| (\vecb r-\vecb r_{n \vecb k}) \times \vecb v_{\vecb k} | u_{n\vecb k} \rangle$, where 
$\vecb{v}_{\vecb k} =  e^{-i \vecb k \cdot \vecb{r} } \ \vecb{v} \ e^{i \vecb k \cdot \vecb{r} }$. 
In fact, any other choice would produce a gauge dependent result, which should not be the case for an observable quantity. Thus,
we write the gauge-covariant expression for the OAM  as 
\begin{align}
\vecb L_{nn'} ( \vecb k) &= \langle \psi_{n \vecb k}| \vecb L|  \psi_{n' \vecb k} \rangle = \frac{m_e}{2} \langle u_{n \vecb k}|(\vecb{r}-\vecb{{\mathcal A}}_{n \vecb k})   \times \vecb{v}_{\vecb k}  \nonumber \\
 & \qquad \qquad \qquad -\vecb{v}_{\vecb k} \times (\vecb{r}-\vecb{{\mathcal A}}_{n'\vecb k})  | u_{n' \vecb k}  \rangle.
 \label{eqm}
\end{align}

To circumvent the ill-defined nature of the position operator for the Bloch functions, we work in the Wannier function representation.
{\Bl Starting from the early calculations for real materials\cite{pawlowska}, the Wannier functions have found multiple uses in the properties of solids such as the electric polarization\cite{vanderbilt_pol}, and they have also
been successfully used to compute the orbital magnetization\cite{vanderbilt_prl, vanderbilt_prb}.
The latter 
involves the band-diagonal elements of the OAM operator, while transport involves both the diagonal and off-diagonal elements.
}


\section{Wannier Basis}

We transform the Bloch states to the Wannier function  basis:
\begin{align}
    | u_{n \vecb k} \rangle &=  \frac{1}{\sqrt N} \sum_{\vecb{R} } e^{i \vecb k \cdot (\vecb R - \vecb r )} \ | n \vecb{R} \rangle, \nonumber \\
    | n \vecb{R} \rangle &=  \frac{ \sqrt{N} ~ \Omega_c }{(2 \pi)^3} ~ \int_{\rm BZ} d \vecb k ~ e^{i \vecb k \cdot ( \vecb{r}  - \vecb{R})} ~ | u_{n \vecb k} \rangle, 
    \label{eq1}
\end{align}
where $| u_{n \vecb k}\rangle$ is the cell-periodic part of the Bloch function, $|\psi_{n\vecb k} \rangle = e^{i\vecb k \cdot \vecb r} | u_{n \vecb k} \rangle$,
$| n \vecb{R} \rangle$ is the Wannier function,  $\vecb R$ being the cell position and $n$ being the band index,
$N$ is the number of unit cells in the crystal, and $\Omega_c$ is the unit cell volume. 
We proceed to evaluate Eq. (\ref{eqm}) in several pieces. The matrix element of $r \times \vecb v_{\vecb k}$ is
\begin{align}
    &  \langle u_{n \vecb k} |  \vecb{r}   \times \vecb{v}_{\vecb k} | u_{n' \vecb k} \rangle \nonumber \\
     & \quad =  \frac{1}{N} \sum_{ \vecb{R} \vecb{R'} }  e^{i \vecb k \cdot (\vecb{R'} - \vecb{R}) } ~
    \langle n \vecb{R} |  ~  (\vecb{r} - \vecb{R} ) \times \vecb{v}    ~| n' \vecb{R'} \rangle \nonumber \\
    & \qquad + \frac{1}{N}\sum_{ \vecb{R}  \vecb{R'} }  e^{i \vecb k \cdot (\vecb{R'} - \vecb{R}) } ~
     \vecb{R}  \times \langle n \vecb{R} |   \vecb{v}    | n' \vecb{R'} \rangle, 
     \label{eq2}
    \end{align}
    which immediately follows from the Wannier function definition, Eq. (\ref{eq1}).
Evaluating the same for $\vecb v_{\vecb k} \times \vecb r$, we find (Supplemental Material \cite{SM}) that the second term in Eq. (\ref{eq2}) cancels out with its counterpart, which is an advantage of using the symmetric form, Eq. (\ref{eq4}).
With this cancellation, the remaining terms are
\begin{align}
    &\langle u_{n \vecb k} | ( \vecb{r}   \times \vecb{v}_{\vecb k} - \vecb{v}_{\vecb k} \times \vecb{r} )  | u_{n' \vecb k} \rangle \nonumber \\
     &= \frac{1}{N}\sum_{ \vecb{R}  \vecb{R'} }  e^{i \vecb k \cdot (\vecb{R'} - \vecb{R}) } ~
    \langle n \vecb{R} |  ~ ( \vecb{r} - \vecb{R} )  \times \vecb{v}    ~| n' \vecb{R'} \rangle \nonumber \\
    & \quad - \frac{1}{N}\sum_{ \vecb{R}  \vecb{R'} }  e^{i \vecb k \cdot (\vecb{R'} - \vecb{R}) } ~\langle n \vecb{R} |  ~  \vecb{v} \times ( \vecb{r} - \vecb{R'} )   ~| n' \vecb{R'} \rangle  \nonumber \\
    &=\sum_{ \vecb{R} }  e^{i \vecb k \cdot \vecb{R} } ~ \langle n 0 |  ~   \vecb{r}   \times  \vecb{v}  ~ | n'  \vecb{R}  \rangle 
    -   e^{-i \vecb k \cdot \vecb{R}  } \langle n \vecb{R}  |  ~  \vecb{v} \times  \vecb{r} ~| n' 0 \rangle
    \label{eq5}
\end{align}
where in obtaining the second equality, we have used the translational symmetry to get rid of one lattice summation.
With the velocity operator 
\begin{align}
    \vecb{v} = -\frac{i}{\hbar}  ~ [ \vecb{r},\mathcal{H} ],
    \label{eq6}
\end{align}
the first term in Eq. (\ref{eq5}) becomes
\begin{align}
    T_1 = \frac{i }{\hbar} \sum_{ \vecb{R} }  e^{i \vecb k \cdot \vecb{R} } ~ \langle n 0 |  ~ \vecb{r} \times \mathcal{H} ~ \vecb{r}   ~ | n'  \vecb{R}  \rangle .
    \label{eq7}
\end{align}

Unlike the case for the Bloch functions, the position operator is well defined for the Wannier functions. 
From its definition, Eq. (\ref{eq1}), we easily find, using periodic gauge for the Bloch functions, the result (Supplemental Material \cite{SM})
\begin{align}
    \vecb{r} ~ | n \vecb{R} \rangle 
     &= \vecb{R} ~ | n \vecb{R} \rangle + \frac{ \sqrt{N} ~ \Omega_c }{(2 \pi)^3} ~ \int_{\rm BZ} d \vecb k ~  e^{i \vecb k \cdot  ( \vecb{r} - \vecb{R} )} ~ | i \partial_{\vecb k} u_{n \vecb k} \rangle ,
     \label{eq8}
\end{align}
so that we have
\begin{align}
    \langle n \vecb R |\vecb  r | n \vecb{R} \rangle = \vecb{R} + \frac{ \Omega_c }{(2 \pi)^3} \int_{\rm BZ} d \vecb k ~ \vecb {\mathcal{A}}_{n \vecb k} .
\end{align}
Note that the center of the Wannier function is offset from the lattice position by the  Berry connection term, 
{\Bl which though it is gauge dependent, the final results will be gauge invariant.}

Turning now to  the matrix element in Eq. (\ref{eq7}), substitution of Eq. (\ref{eq8}) leads to
\begin{align}
    & \langle n 0 |  ~ \vecb{r} \times \mathcal{H} ~ \vecb{r}   ~ | n'  \vecb{R}  \rangle \nonumber \\
    &= -i\  \bigg( \frac{ \sqrt{N} ~ \Omega_c }{(2 \pi)^3} \bigg)^2 \iint_{\rm BZ} d \vecb k' ~ d \vecb k'' ~ e^{-i \vecb k' \cdot \vecb{R} } \nonumber \\
    & ~ \langle \partial_{\vecb k''} u_{n \vecb k''} | \times e^{-i \vecb k'' \cdot \vecb{r}} \mathcal{H} e^{i \vecb k' \cdot \vecb{r}} ~ ( \vecb{R}  \ | u_{n' \vecb k'} \rangle + | i \ \partial_{\vecb k'} u_{n' \vecb k'} \rangle ).
    \label{eq9}
\end{align}
One momentum integral in Eq. (\ref{eq9}) can be removed because the integrands are $\delta$ functions, e. g.,  
\begin{align}
   & \langle \partial_{\vecb k''} u_{n \vecb k''} |  e^{-i \vecb k'' \cdot \vecb{r}} \mathcal{H} e^{i \vecb k' \cdot \vecb{r}}  | u_{n' \vecb k'} \rangle \nonumber \\
    & = \frac{(2 \pi)^3}{N \Omega_c} ~ \delta(\vecb k' - \vecb k'') ~\varepsilon_{n' \vecb k'} ~ \langle \partial_{\vecb k'} u_{n \vecb k'} |  u_{n' \vecb k'} \rangle,
\end{align}
while the second one is removed by performing the lattice summation  in Eq. (\ref{eq7}) (Supplemental Material \cite{SM}). The result is
\begin{align}
T_1 
    &=   \frac{ \Omega_c}{\hbar  (2 \pi)^3} \int\limits_{\rm BZ} d \vecb k' ~ \sum_{\vecb{R}} e^{i \vecb{R} \cdot (\vecb k - \vecb k' )}  
      \big[ \varepsilon_{n' \vecb k'}  \langle \partial_{\vecb k'} u_{n \vecb k'}    | u_{n' \vecb k'} \rangle \times \vecb{R}  \nonumber \\
      &\hspace{3cm}  + \langle \partial_{\vecb k'} u_{n \vecb k'} | \times \mathcal{H}_{\vecb k'}  | i \partial_{\vecb k'} u_{n' \vecb k'} \rangle \big] \nonumber \\ 
     &= \frac{i }{\hbar}  \big[ \langle \partial_{\vecb k} u_{n \vecb k} | \times \mathcal{H}_{\vecb k}  | \partial_{\vecb k} u_{n' \vecb k} \rangle  +\partial_{\vecb k}  \times   ( \varepsilon_{n' \vecb k}  \langle \partial_{\vecb k} u_{n \vecb k}  | u_{n' \vecb k} \rangle ) \big].
     \label{eq12}
\end{align}
Following the same algebra, we find the the second term in Eq. (\ref{eq5}) to be
\begin{align}
     &T_2= \frac{i }{\hbar} \sum_{ \vecb{R} }  e^{-i \vecb k \cdot \vecb{R}  } ~\langle n \vecb{R}  |  ~  \vecb{r} ~ \mathcal{H} \times  \vecb{r} ~| n' 0 \rangle 
     \nonumber \\
     &=  \frac{i }{\hbar} \big[ \langle \partial_{\vecb k} u_{n \vecb k} | \times \mathcal{H}_{\vecb k} | \partial_{\vecb k} u_{n' \vecb k} \rangle - \partial_{\vecb k} \times  \left( \varepsilon_{n \vecb k}  \langle u_{n \vecb k} |  \partial_{\vecb k} u_{n' \vecb k} \rangle \right) \big].
     \label{eq13}
\end{align}
Substituting $T_1$ and $T_2$ into  Eq. (\ref{eq5}), we get
\begin{align}
   &\langle u_{n \vecb k} | ( \vecb{r}   \times \vecb{v}_{\vecb k} - \vecb{v}_{\vecb k} \times \vecb{r} )  | u_{n' \vecb k} \rangle \nonumber \\
     &=  \hbar^{-1} 
     \big[ \ 2 i \ \langle \partial_{\vecb k} u_{n \vecb k} | \times \ \mathcal{H}_{\vecb k} \ | \partial_{\vecb k} u_{n' \vecb k} \rangle \nonumber \\
     & \quad - \partial_{\vecb k} \times \big\{ \big( \varepsilon_{n \vecb k} \ + \varepsilon_{n' \vecb k} \big) ~ \vecb{\mathcal{A}}_{nn' \vecb k}   \big\}  \big],
     \label{eq14}
\end{align}
where $\vecb{{\mathcal A}}_{nn'} (\vecb k) = i \langle  u_{n \vecb k}   | \partial_{\vecb k} u_{n' \vecb k} \rangle $. 

\begin{widetext}
The remaining term in Eq. (\ref{eqm}) is easily evaluated:
\begin{align}
   &\frac{m_e}{2} \langle u_{n \vecb k} | ( - \vecb{\mathcal{A}}_{n\vecb k} \times \vecb{v}_{\vecb k} \ + \ \vecb{v}_{\vecb k} \times \vecb{\mathcal{A}}_{n'\vecb k} ) | u_{n' \vecb k} \rangle   = -\frac{m_e}{2}  (\vecb{\mathcal{A}}_{n\vecb k} + \vecb{\mathcal{A}}_{n'\vecb k}) \times \vecb{v}_{nn' \vecb k},
   \label{eq15}
\end{align}
where $\vecb{v}_{nn'\vecb k} = \langle u_{n \vecb k} | \vecb v_k | u_{n' \vecb k} \rangle = \hbar^{-1} \langle u_{n\vecb{k}} | \partial_{\vecb{k}} \mathcal{H}_{\vecb{k}} | u_{n'\vecb{k}} \rangle$. 
Adding up Eqs. (\ref{eq14}-\ref{eq15}),  using the equality $\partial_{\vecb k} \times \vecb{\mathcal{A}}_{nn'}(\vecb k) = i \langle \partial_{\vecb k} u_{n\vecb k} | \times | \partial_{\vecb k} u_{n'\vecb k} \rangle$, and noting that
$ \vecb{\mathcal{A}}_{n\vecb k} \equiv \vecb{\mathcal{A}}_{nn\vecb k}$, we get the desired answer  
    \begin{subequations}
        \begin{align}
            \vecb{L}_{nn'} (\vecb k) &= \frac{im_e }{\hbar}  \ \langle \partial_{\vecb k} u_{n \vecb k} | \times  \big( \mathcal{H}_{\vecb k} - \frac{ \varepsilon_{n \vecb k} \ + \varepsilon_{n' \vecb k} }{2} \big) \ | \partial_{\vecb k} u_{n' \vecb k} \rangle 
            +\frac{m_e}{2} \big[ \vecb{\mathcal{A}}_{nn' \vecb k} \times ( \vecb{v}_{nn \vecb k} + \vecb{v}_{n' n' \vecb k} )  - (\vecb{\mathcal{A}}_{n\vecb k} + \vecb{\mathcal{A}}_{n'\vecb k}) \times \vecb{v}_{nn' \vecb k} \big], 
\label{eq16a} \\
            &= \frac{im_e\hbar}{2} \big[ \sum_{m \neq n, n'} \big(
            \frac{1}{\varepsilon_{m \vecb k} - \varepsilon_{n \vecb k}} +\frac{1}{\varepsilon_{m \vecb k} - \varepsilon_{n' \vecb k}}\big) \ \vecb{v}_{nm \vecb k} \times \vecb{v}_{mn' \vecb k}  + \frac{(\vecb{v}_{nn, \vecb k} + \vecb{v}_{n'n' \vecb k}) \times \vecb{v}_{nn' \vecb k}}{\varepsilon_{n \vecb k} - \varepsilon_{n' \vecb k}} \big],
\label{eq16b}
        \end{align} 
\label{eq16}
    \end{subequations}
\end{widetext}
where the sum in Eq. (\ref{eq16b}) is over the first term only. 
The second line follows after some straightforward algebra  (Supplemental Material \cite{SM}) using the well-known result for the Bloch function derivative\cite{Vanderbilt-Book}
\begin{equation}
    |\partial_{\vecb k} u_{n \vecb k}\rangle = -i \vecb{\mathcal{A}}_{n \vecb k} |u_{n \vecb k} \rangle + \sum_{m \neq n} 
    \frac{\langle u_{m \vecb k} | \partial_{\vecb k} {\mathcal H}_{\vecb k}  | u_{n \vecb k} \rangle}{\varepsilon_{n \vecb k} - \varepsilon_{m \vecb k}} |u_{m \vecb k} \rangle.
\end{equation}

The OAM expression Eq. (\ref{eq16}) is the central result of this paper, and it is evidently gauge covariant.
 It is easy to show that (Supplemental Material \cite{SM}) Eq. (\ref{eq16}) is consistent with the well-known conditions under time-reversal ($\mathcal{ T}$) and inversion ($\mathcal{ I}$) symmetry, viz., 
\begin{align}
    \vecb L_{nn'} (-\vecb k) &= \vecb L_{nn'} ( \vecb k) \qquad &({\rm  \mathcal{ I}\  present)}, \nonumber \\
    \vecb L_{nn'} (-\vecb k) &= - \vecb L_{nn'}^{*} ( \vecb k) \qquad &({\rm  \mathcal{ T}\  present)}.
    \label{eq18}
\end{align}
Note that the band-diagonal term, $n = n'$, in Eq. (\ref{eq16}) agrees with the well-known results for the OAM in the literature\cite{niu_1996, niu_1999, david_2008}. The off-diagonal term, which enters in the calculation of the OHC using the Kubo expression,  has been derived by several authors using the Bloch function basis\cite{ohe_vignale, ohe_pezo, ohe_rappoport} with differing results, which indicates the subtlety of computing the position matrix elements for the Bloch functions.

\section{Haldane Model}

To validate our results, we have calculated the OAM directly from its definition, Eq. (\ref{eq4}), taking the example of the Haldane model as discussed below. The complete agreement between these two independent calculations establishes the validity of our OAM expression.
\begin{figure}
    \centering
    \includegraphics[scale=0.24]{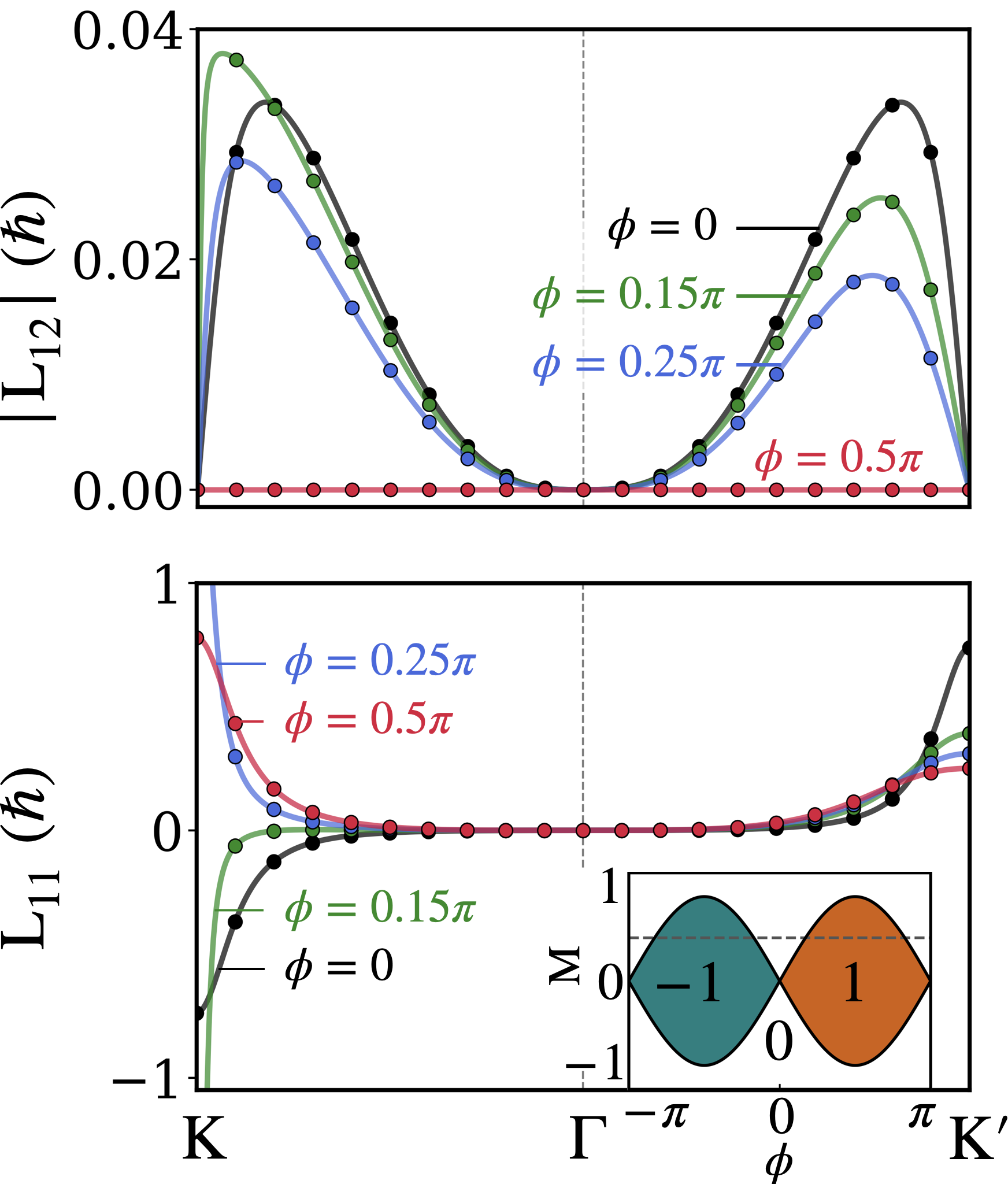}
    \caption{OAM matrix elements for the Haldane model from two different calculations: (i) From the OAM expression Eq. (\ref{eq16}) (solid lines) and (ii) From the direct evaluation of OAM for the $60 \times 60$ finite lattice (circles).
The off-diagonal matrix element $L_{12} > 0$ along $\Gamma-K$ and $< 0$ along $\Gamma-K'$, and only the magnitude is shown in the top figure for convenience.     
{\Bl
The {\it inset} in the bottom figure indicates the Chern number for the Haldane model ($ C = 0$ or $\pm 1$) and the dashed line corresponds to the parameters used in the calculation with four different values of $\phi = (0, 0.15, 0.25, 0.5) \ \pi$.}
}
\label{fig2}
\end{figure}

In the first method, we evaluate the OAM matrix elements 
using the bulk expression Eq. (\ref{eq16}).
In the second method, we compute the matrix elements of the OAM operator directly 
from its definition in real space, Eq. (\ref{eq4}),
by taking a finite lattice of size $N \times N$ ($N = 60$) with the periodic boundary condition (PBC). 
{\Bl
In the finite-lattice calculation, the wave functions for a discrete set of momentum points in the Brillouin zone  are accessible, which are identical to the infinite-lattice wave functions. Therefore,  the OAM matrix elements can be computed exactly using the finite lattice with PBC at these momentum points, which can then be compared with the results of our OAM expression, Eq. (\ref{eq16}).
}

The Haldane model\cite{Haldane1988} on the honeycomb lattice is
\begin{align}
\mathcal{H} = M \sum_{i} \xi_i c_i^\dagger c_i
+ t_1 \sum_{\langle i,j \rangle} c_i^\dagger c_j
+ t_2 \sum_{\langle i,j \rangle^\prime} e^{i \nu_{ij} \phi} c_i^\dagger c_j ,
\end{align}
where $c_i^\dagger$ and $c_i$ are  spinless fermion creation and annihilation operators for the lattice site $i$, $\langle i,j \rangle$ and $\langle i,j \rangle^\prime$ indicate the nearest-neighbor (NN) and the second-neighbor (2NN) hoppings, respectively, $\xi_i = \pm 1$ introduces a staggered sublattice potential, $M$ being  the ``Semenoff mass," and the factor $\nu_{ij} = \pm 1$ introduces a chirality for the 2NN hopping (clockwise or anticlockwise). 
We take the parameters 
$ t_1 = 1$, $t_2 = 0.15$, and $M = 0.4$, so that by varying the 
complex phase $\phi$, one obtains both the trivial and the topological phases\cite{Haldane1988}, characterized by the Chern number,
{\Bl as indicated in the {\it inset} of the Fig. \ref{fig2}, bottom.
}

To check the accuracy of our finite-lattice calculations, we first computed the Chern number
($ \sigma_{xy} = e^2/h\ C $, $ \sigma_{xy}$ being the Hall conductivity)
for the lowest band in the Haldane model, which involves only the velocity matrix elements and no position or OAM matrix elements. 
The complete agreement (Supplemental Material \cite{SM}) with the well-known results, $C = 0$ or $\pm 1$, depending on the Hamiltonian parameters, validates the accuracy of the finite-lattice calculations.

\begin{figure}
    \centering
    \includegraphics[scale=0.35]{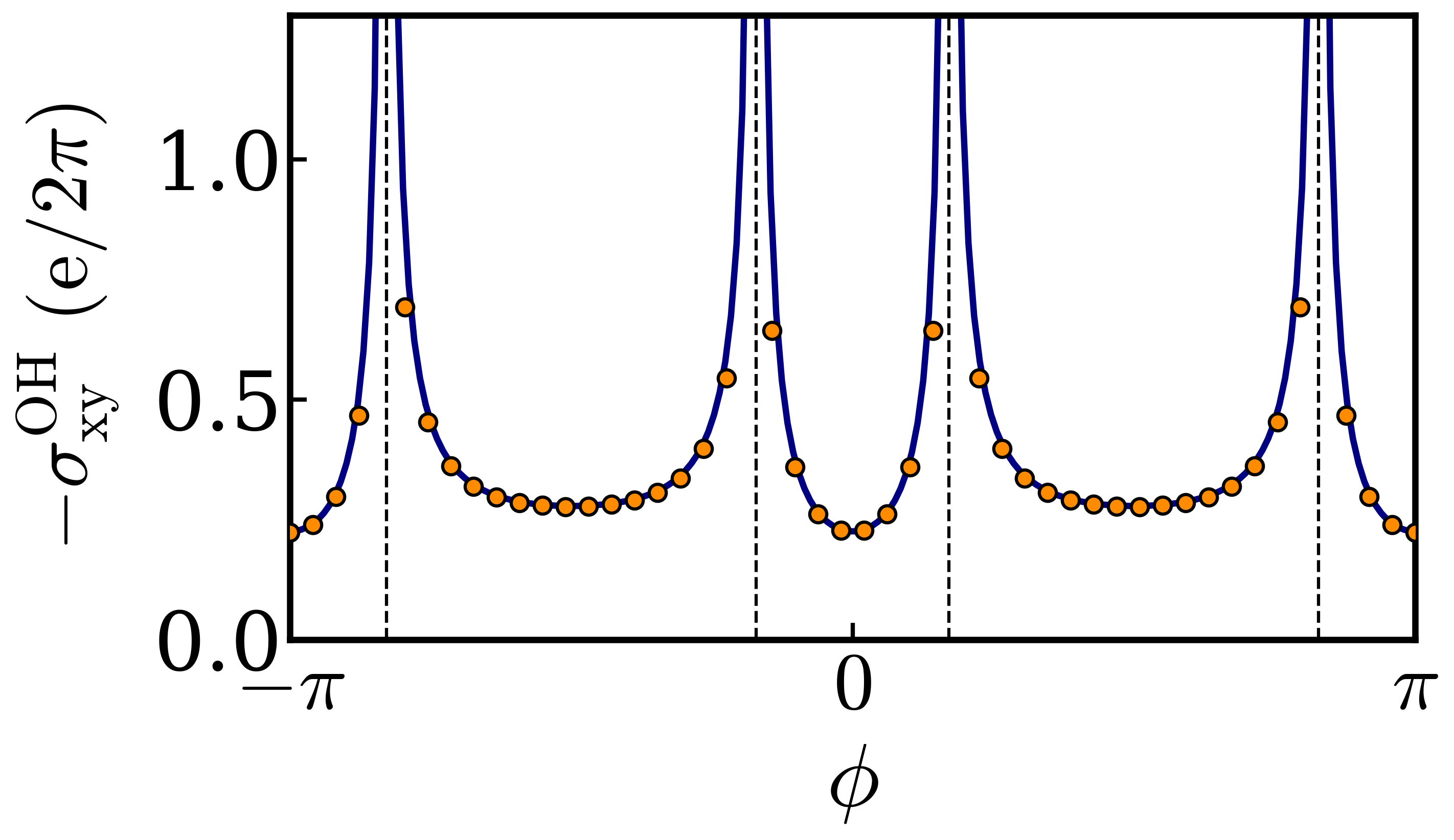}
    \caption{Orbital Hall conductivity in the Haldane model. Full line is obtained from the OAM expression, Eq. (\ref{eq16b}), while the circles are from the direct evaluation of OAM in the finite-lattice model\cite{Note3}. Vertical dashed lines indicate the topological phase boundaries in the Haldane model, where the Chern number changes. 
    }
 \label{fig3}
\end{figure}

Turning now to the OAM matrix elements   for the finite lattice, they are obtained directly from its definition, Eqs. (\ref{eq4}-\ref{eqm}).
In the tight-binding basis $|i\rangle$, the velocity matrix elements are 
\begin{equation}
    \langle i | {\vecb{v}} | j \rangle = \frac{1}{i\hbar} \langle i | ({\vecb{r}}{\mathcal{H}} - {\mathcal{H}}{\vecb{r}}) | j \rangle
    = \frac{1}{i\hbar} (\vecb{r}_i - \vecb{r}_j)\  t_{ij},
    \label{eq20}
\end{equation}
where $t_{ij} = \langle i | \mathcal{H} | j \rangle$ 
is the hopping amplitude and $\vecb{r}_i$ is the position of the $i$-th site.
For the eigenstates of the finite lattice $|\psi_n \rangle$, the off-diagonal elements of the position operator are easily computed from the definition of $\vecb v$, which reads
\begin{equation}
    \langle \psi_n | {\vecb{r}} | \psi_m \rangle = \frac{-i\hbar \langle \psi_n | {\vecb{v}} | \psi_m \rangle} {\varepsilon_n - \varepsilon_m},
    \label{eq21}
\end{equation}
where  the matrix elements $\langle \psi_n | {\vecb{v}} | \psi_m \rangle$   are computed from the eigenfunctions 
$\langle i | \psi_n\rangle$ and Eq. (\ref{eq20}).
On the other hand, the diagonal elements of $\vecb r$ are ill-defined.
However, as shown during the derivation of Eq. (\ref{eq16}), the terms containing these diagonal elements in Eq. (\ref{eq4}) disappear when Eq. (\ref{eqm}) is adopted (Supplemental Material \cite{SM}).
%
Using Eqs. (\ref{eq20} - \ref{eq21}), it is then straightforward to evaluate the OAM matrix elements.

\begin{figure} [h]
\vspace{0.3cm}
    \centering
    \includegraphics[scale=0.30]{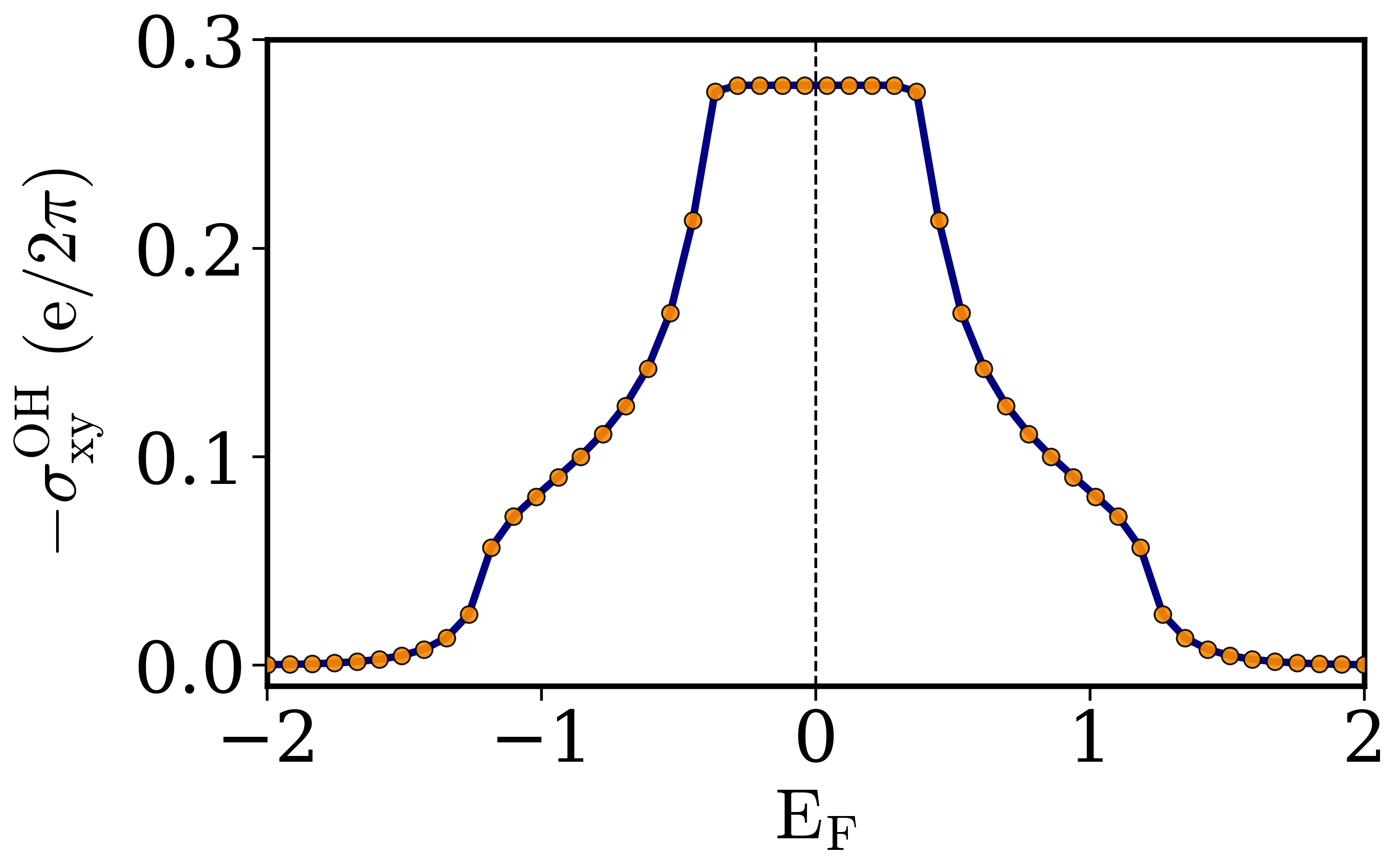}
    \caption{OHC in the Haldane model as a function of the Fermi energy: Results from the expression using Eq. (\ref{eq16b})  vs. the direct  $60 \times 60$ finite-lattice calculation are shown. 
    Parameters are the same as in Fig. 1, except that the phase factor is taken as $\phi = \pi/2$.
    }
\label{fig4}
\end{figure}

Fig. \ref{fig2} shows the key comparison of the paper, where we show the OAM matrix elements computed from the two independent ways.
The complete agreement between the two results establishes the correctness of Eq. (\ref{eq16}). 
%
%
The computed matrix elements follow the required symmetry for OAM, Eq. (\ref{eq18}).
Fig. \ref{fig2} (bottom part) shows that
for $\phi = 0$, where $\mathcal{T}$ is present, the band-diagonal term $L_{11}$ changes  sign when $\vecb k$ is flipped: $L_{11} (\vecb k) = - L_{11} (-\vecb k)$,
while no such symmetry exists for $\phi \ne 0$ due to the broken  $\mathcal{T}$ symmetry. 
As can be inferred from 
Eq. (\ref{eq16b}), the two band-diagonal elements are the same: $\vecb L_{11} = \vecb L_{22}$ for any two-band model such as the Haldane model.
Note that the inversion symmetry is always broken in the Haldane model, unless $M =0$.
Similarly, the off-diagonal elements obey $L_{12} (\vecb k) = -L_{12} (\vecb {-k})$, since   $L_{12}$ turns out to be real for the Haldane model.

{\Bl 
The choice of the Haldane model provides a particularly transparent setting in which to distinguish our formulation from the commonly used approaches. First, within the atom-centered approximation (ACA), the orbital moment vanishes identically for any s-orbital system (like the Haldane model or gapped graphene), while both Eq. (16) and the finite-lattice results yield nonzero diagonal matrix elements for the OAM operator. Furthermore, the OAM expression of Ref. \cite{ohe_pezo}, which is the same as the first term in Eq. (16b), leads to an identically zero  off-diagonal OAM matrix element for any two-band system. The direct numerical calculations of the Haldane model show that this is not true in general, and Eq. (16) correctly reproduces the off-diagonal matrix elements as well.
}

{\Bl
Our result, Eq. (\ref{eq16}), differs from Ref. ~\cite{ohe_pezo}, but agrees with the expression of Rappoport and coworkers ~\cite{ohe_rappoport}, both derived in the Bloch basis.
Apart from deriving the equations from a completely different approach,
we have, importantly, established the correctness of Eq. (16) from the  exact calculations of the Haldane model using the original definition, Eq. (\ref{eq4}), for $\bs L$.
}

\section{Orbital Hall Conductivity}

We now turn to the calculation of the OHC from the Kubo linear response formula, which involves the OAM matrix elements. The Kubo formula is
\begin{equation}
    \sigma_{xy}^{\mathrm{OH}} = \frac{2 e \hbar}{A} \ \mathrm{Im} \sum_{n}^{\mathrm{occ}} \sum_{m \neq n} \frac{ \langle \psi_n | J_x  | \psi_m \rangle \langle \psi_m | {v}_y | \psi_n \rangle}{(\varepsilon_n - \varepsilon_m)^2},
    \label{eq19}
\end{equation}
where $J_x= 2^{-1} \left({v}_x {L}_z + {L}_z {v}_x \right)$ is the orbital-current operator.
For the Hall conductivity $\sigma_{xy}$, ${J}_x$ is simply replaced by $-e v_x$.
The computed OHC from the two methods
are shown in 
Fig. \ref{fig3}, which agree with each other. 
The divergence in the OHC at the topological crossover points seen in Fig. \ref{fig3} 
also occurs for the orbital moment 
{\Bl 
due to the gap closing between the two bands. Such divergences have been earlier noticed for the Dirac points in gapped graphene as well\cite{ohe_vignale}.
}
In Fig. \ref{fig4}, we plot the OHC as a function of the Fermi energy,
which also shows excellent agreement between the two methods.

\section{Conclusion}

In conclusion, by working in the Wannier function representation, we derived an expression for the OAM for the crystalline solid.
The correctness of our formulation was validated by computing the corresponding quantities for the finite-lattice Haldane model, where the OAM was calculated directly from its definition. Our results are valid for both metals and insulators, which was illustrated for the Haldane model as well.

\end{document}